\documentclass[10pt,letterpaper,twocolumn]{article} %% two column, final layout

\usepackage{ol2}
\usepackage[draft]{hyperref}
\usepackage{amsmath}

\begin{document}

\twocolumn[ %% activate for two-column option

\title{Buffer gas-assisted polarization spectroscopy of $^6$Li }

%% For REVTeX it is possible to automate superscript and e-mail callouts with the superscriptaddress option; see REVTeX4 documentation.

\author{Nozomi Ohtsubo, Takatoshi Aoki, and Yoshio Torii}

\address{
Institute of Physics, The University of Tokyo, Komaba \\ 3-8-1 Komaba, Megro-ku, Tokyo 153-8902, Japan
\\
}

\begin{abstract}
\noindent
We report on the demonstration of Doppler-free polarization spectroscopy of the D2 line of $^6$Li atoms. Counterintuitively, the presence of an Ar buffer gas, in a certain pressure range, causes a drastic enhancement of the polarization rotation signal. The observed dependence of the signal amplitude on the Ar buffer pressure and the pump laser power is reproduced by calculations based on simple rate equations. We performed stable laser frequency locking using a dispersion signal obtained by polarization spectroscopy for laser cooling of $^6$Li atoms.
\end{abstract}


 ]%% activate for two-column option

\noindent 
Since the first realization of Bose-Einstein condensation in a dilute atomic gas, extensive studies of quantum-degenerate gases have been carried out \cite{Review}. Recently, there has been increasing interest in quantum-degenerate fermions as a tool to explore the physics of strongly-correlated electron systems \cite{Ketterle2008}. Lithium-6 is one of the workhorses for the study of quantum-degenerate fermions since the strength of the inter-atomic interaction is widely tunable via a broad Feshbach resonance \cite{Dieckmann}. Lithium-6 is also attractive for the study of ultracold hetero-nuclear molecules since the molecules composed of lithium, such as LiRb and LiCs, have relatively large electric dipole moments \cite{dipole-moment}, which are beneficial to explore novel quantum phases of ultracold molecules \cite{quantum -phase}. We note that the natural abundance of $^6$Li is relatively large ($\sim$ 8$\%$) and enriched samples are easily available as compared with the other alkali or alkali-earth atoms.

To date, quantum-degenerate samples of fermions have been produced by the combination of laser cooling and evaporative cooling \cite{Ketterle2008}. For laser cooling, it is necessary to lock the frequency of the cooling laser to a specific resonance line. Frequency modulation (FM) spectroscopy is widely used to obtain an error signal for laser frequency locking \cite{Bjorklund}. Alternatively, modulation-free frequency stabilization techniques, such as dichroic atomic vapor laser lock (DAVLL) \cite{DAVLL}, have been employed. Doppler-free polarization spectroscopy (DPS) \cite{Wieman1976} can also offer a dispersive error signal at the atomic resonances, and has been used for modulation-free laser frequency locking to the rubidium D2 line \cite{Yoshikawa, Pearman }. 

Surprisingly, there has been no report on DPS of lithium atoms in the literature so far. In this Letter, we demonstrated DPS of the D2 line of $^6$Li atoms and proved that DPS provides dispersion signals suitable for laser frequency locking. We studied the effect of an Ar buffer gas on the polarization spectrum, and found that the amplitude of the dispersion signal at the cooling transition is drastically increased by the help of the Ar buffer gas. The observed dependence of the signal amplitude on the Ar buffer pressure and the pump laser power was reproduced by calculations based on simple rate equations considering all the relevant magnetic sublevels. We also demonstrated laser cooling of $^6$Li atoms with a laser which was frequency stabilized using a DPS dispersion signal enhanced by the Ar buffer gas.
 
\begin{figure}[b]
\centerline{\includegraphics[width=8.5cm]{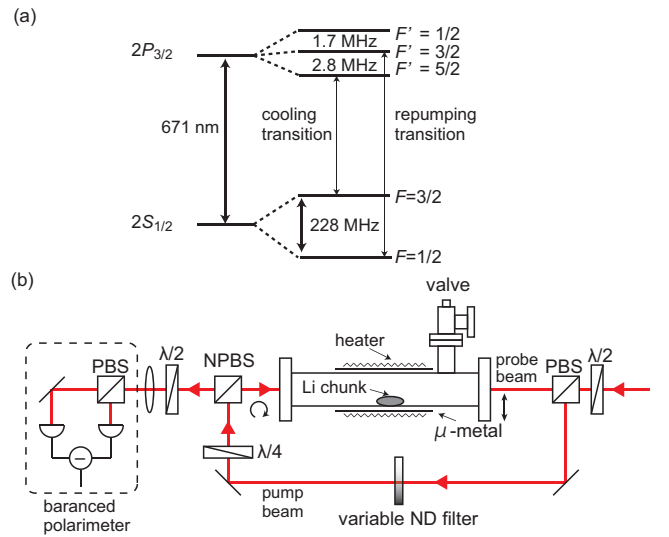}}
\caption{(a) Relevant $^6$Li energy levels. (b) Schematic diagram of the experimental setup. PBS, polarization beam splitter; NPBS, non-polarization beam splitter; $\lambda$/2, half-wave plate; $\lambda$/4, quarter-wave plate. }
\end{figure}
\begin{figure*}[htb]
\centerline{\includegraphics[width=17cm]{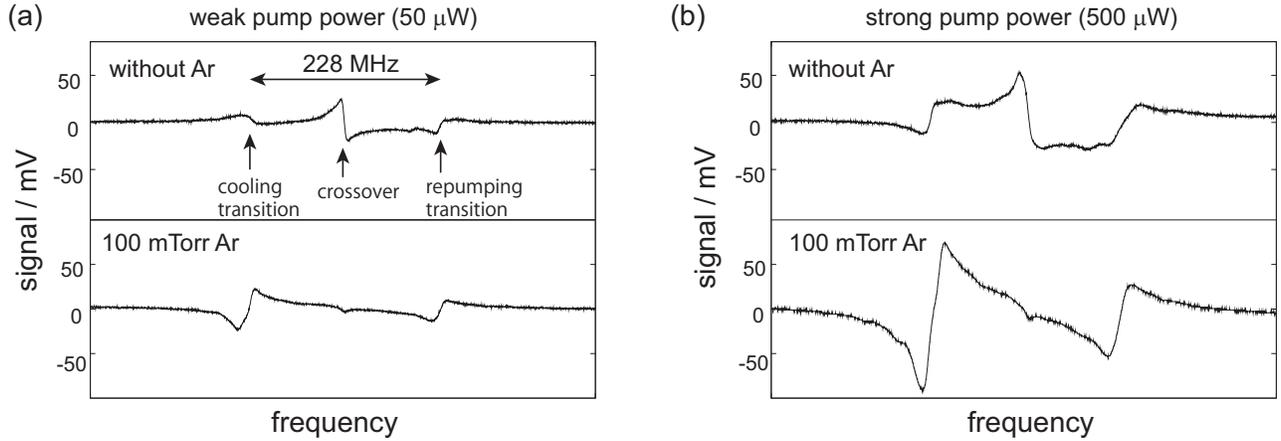}}
\caption{Doppler-free polarization spectra of the 2S$_{1/2} \rightarrow$2P$_{3/2}$ transitions of $^6$Li at zero and 100-mTorr Ar buffer gas pressures for (a) a weak (50 $\mu$W) and (b) a strong (500 $\mu$W) pump power. }
\end{figure*}
\begin{figure}[b]
\centerline{\includegraphics[width=8.5cm]{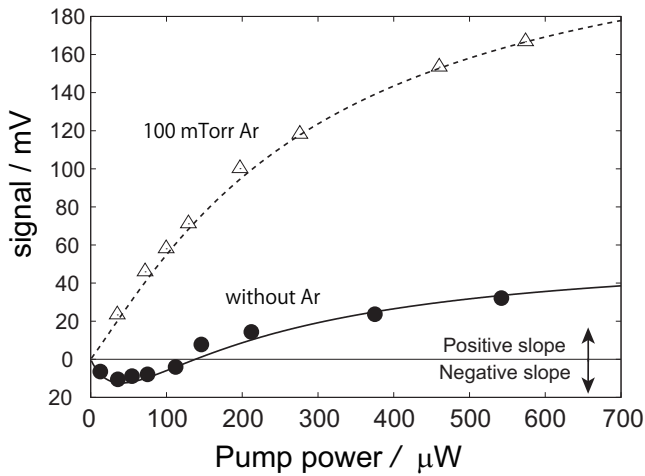}}
\caption{Dependence of the dispersion signal at the cooling transition on the pump power. The vertical axis represents the peak to peak amplitude of the dispersion signal. The open triangles and the filled circles represent the experimental data at zero and 100-mTorr Ar buffer gas pressures, respectively. The solid line and the dashed line show calculations for each condition.}
\end{figure}
Figure 1(a) shows the energy diagram of $^6$Li. For laser cooling of $^6$Li atoms, the cooling laser is tuned to the 2S$_{1/2}$, $F=3/2$ $\rightarrow$ 2P$_{3/2}$, $F^{\prime}$=5/2 transition, whereas the repumping laser is tune to the 2S$_{1/2}$, $F=1/2$ $\rightarrow$ 2P$_{3/2}$, $F^{\prime}$=3/2 transition. The D2 line of $^6$Li is unique in that the hyperfine-structure splittings of the 2P$_{3/2}$ states are smaller than the natural linewidth of 5.9 MHz, therefore the atoms in the upper (2S$_{1/2}$, $F=3/2$) hyperfine level are easily pumped to the lower (2S$_{1/2}$, $F=1/2$) hyperfine level by the cooling laser via a few absorption- spontaneous emission cycles (hyperfine pumping). 

Figure 1(b) shows the experimental setup. A 5-g chunk of $^6$Li (enriched $>$ 95$\%$) was installed in the middle of a 50-cm-long stainless tube ended by two ICF70 glass viewports. The tube had a valve to introduce an Ar buffer gas with a desirable pressure ranging from 0 to 100 mTorr. The central part ($\sim$ 20 cm) of the tube was magnetically shielded by winding a sheet of $\mu$-metal and heated at 350 $^\circ$C by a tape heater. A 671-nm laser beam was derived from an external-cavity diode laser (ECDL) using an AR-coated diode laser (Eagleyard EYP-RWE-0670-00703). 

The optical setup for DPS of $^6$Li atoms was basically the same as Ref.\cite{Yoshikawa}.  A circularly-polarized pump beam and a linearly-polarized probe beam, both of which had almost the same diameters ($\sim$ 2 mm), were sent to the vapor cell in a Doppler-free configuration. The power of the pump beam was varied from 0 to 500 $\mu$W using a neutral density filter, whereas that of the probe beam was fixed at 50 $\mu$W (a power of $\sim$150 $\mu$W corresponds to the saturation intensity of 2.5 mW/cm$^2$). Circular birefringence induced by the pump beam was monitored by the probe beam as a rotation of the polarization axis. The angle of polarization rotation was then converted to the electronic signal using a balanced polarimeter, which consisted of a balanced photo detector and a polarizing beamsplitter (PBS) \cite{Yoshikawa}. 

We performed DPS with various pump powers and Ar buffer pressures. Figure 2 shows typical polarization spectra at zero and 100-mTorr Ar buffer pressures for a weak (50 $\mu$W) and a strong (500 $\mu$W) pump power. We observed three dispersion signals at the frequencies of the cooling, the crossover, and the repumping transitions as indicated by the arrows in the figure. The 2P$_{3/2}$ hyperfine structure were not resolved. Each dispersion signal displayed different behavior as we introduced an Ar buffer gas. The amplitude of the dispersion signal at the cooling (crossover) transition increased (decreased) with increasing the pressure of the Ar buffer gas. Note that, for a weak pump power, the slope of the dispersion signal at the cooling transition changed its sign, which indicates that there are different physical origins of circular birefringence as discussed below. 

Figure 3 shows the dependence of the amplitude of the dispersion signal at the cooling transition on the pump power. The amplitudes of the dispersion signal with positive (negative) slopes are plotted in the positive (negative) direction of the vertical axis.  Without an Ar buffer gas, the slope changes its sign at the pump power around 150 $\mu$W, and then the amplitude increases monotonically with increasing the pump power. With a 100-mTorr Ar buffer gas, the amplitudes are much larger than those without an Ar buffer gas for any pump power. Taking the pump power of 500 $\mu$W, for example, the amplitude with a 100-mTorr Ar buffer gas is five times the amplitude without an Ar buffer gas (see Fig. 2(b)). 

We compared the experimental data with calculations based on simple rate equations considering all the relevant magnetic sublevels as explored in Ref.\cite{VCC}, where the effect of velocity-changing collisions (VCCs) caused by the Ar buffer gas is included as a thermal relaxation of the velocity distribution for each magnet sublevel (VCCs remove atoms from the zero velocity group out of resonance and bring atoms from other velocity group into resonance). We also included the effect of hyperfine pumping due to the probe beam itself in the rate equations. The relaxation rate due to VCCs was set to the typical value of 10 MHz/Torr \cite{Budker}. The optical density of the $^6$Li gas and the diameter of the pump beam were chosen to fit the experimental data. The experimental data are in good agreement with the calculations assuming a pump beam diameter of 2.5 mm (and a resultant transit-time broadening of 90 kHz), which is close to the actual pump beam diameter of $\sim$ 2 mm.

Without an Ar buffer gas, the typical amplitude of the dispersion signal observed at the cooling transition corresponds to the polarization rotation angle of $\sim$ $10^{-3}$ rad. This rotation angle is about an order of magnitude smaller than those observed for Rb and Cs vapors with comparable optical densities \cite{Hughes2006}. For the D2 lines of Rb and Cs, the hyperfine-structure splittings of the excited states ($n$P$_{3/2}$) are much larger than the natural linewidth and the cooling transition is nearly closed. Therefore, the atoms are easily spin-polarized by a relatively weak circularly-polarized pump beam via a few absorption-spontaneous emission cycles, and exhibit circular birefringence \cite{Hughes2006}. We call this birefringence due to spin-polarization as Type-I. On the other hand, as mentioned above, $^6$Li atoms are easily pumped to the lower hyperfine state by a circularly-polarized pump beam, leading to a reduced efficacy of spin-polarization. If we increase the pump power, another type of circular birefringence emerges: the $\sigma^{+}$ ($\sigma^{-}$) pump beam saturates the $\sigma^{+}$ ($\sigma^{-}$) transitions and weakens the strength of the interaction with the $\sigma^{+}$ ($\sigma^{-}$) component of the linearly-polarized probe beam, resulting in circular birefringence. We call this saturation-induced birefringence as Type-II \cite{Wieman1976}. Unfortunately, the effects of these two types of birefringence cancel out for the cooling transition \cite{endnote}, resulting in a small dispersion signal for any pump power. The change of the sign seen in Fig.3 is explained by a slight difference between the dependences of the two types of birefringence on the pump power.

The enhancement of the dispersion signal at the cooling transition by introducing an Ar buffer gas can be qualitatively explained as follows. VCCs due to the Ar buffer gas prevent spin-polarization of the $^6$Li atoms, leading to a suppression of Type-I birefringence. On the other hand, Type-II birefringence is not so affected by the presence of the Ar buffer gas since the population relaxation rate due to VCCs at Ar buffer pressures below $\sim$ 100 mTorr is much smaller than the spontaneous decay rate \cite{VCC}. As a consequence, the unfortunate cancellation of the two types of birefringence is broken and Type-II birefringence manifests itself. Moreover, the reduction of the population of the zero velocity group due to hyperfine pumping by the probe beam is compensated to some extent by the help of VCCs. As a result, the dispersion signal at the cooling transition is significantly enhanced by the Ar buffer gas. 

The origin of the crossover signal between ground-state hyperfine levels is a bump in the velocity distribution created by hyperfine pumping due to the pump beam. VCCs due to the Ar buffer gas, as mentioned above, smear the bump, therefore the introduction of the Ar buffer gas leads to a diminishment of the crossover signal as seen in Fig. 2.

We performed laser frequency locking using a dispersion signal obtained by polarization spectroscopy for laser cooling of $^6$Li atoms. The laser was locked with a stability of less than 1 MHz for a day. Using this frequency stabilized laser, we demonstrated magneto-optical trapping of $^6$Li atoms.

In conclusion, we have demonstrated Doppler-free polarization spectroscopy of the D2 line of $^6$Li atoms and studied the effect of an Ar buffer gas. We found that the dispersion signal at the cooling transition is drastically enhanced by an Ar buffer gas, which is explained by two effects of velocity-changing collisions with Ar atoms: the breaking of cancellation between two types of birefringence and the compensation of hyperfine pumping by the probe beam. We performed stable laser frequency locking using a dispersion signal obtained by polarization spectroscopy for magneto-optical trapping of $^6$Li atoms.% For a long-term operation of the Li vapor cell, introduction of a buffer gas is necessary to protect the viewports from Li atoms, which strongly react with glass. Our work shows another benefit of introducing a buffer gas in the Li vapor cell.

We thank Daisuke Ikoma for technical assistance in the development of the laser system. This work is supported by MEXT (KAKENHI 22104501).

%\bibliographystyle{ol}

%\bibliography{test}

\begin{thebibliography}{99}
\bibitem{endnote}
For the cooling transition, the atoms which are spin-polarized by the $\sigma^{+}$ ($\sigma^{-}$) pump beam interact predominantly with the $\sigma^{+}$ ($\sigma^{-}$) component of the probe beam if the saturation effect due to the pump beam is neglected. See, for example, Ref.\cite{Yoshikawa}.
\end{thebibliography}

\end{document}